\documentclass[a4paper,11pt]{article}
\usepackage{jinstpub}

\usepackage{subcaption}
\usepackage{siunitx}
\usepackage[version=4]{mhchem}

\proceeding{\ 8\textsuperscript{th} International Conference on Micro Pattern Gaseous Detectors - MPGD2024\\
14-18 October 2024\\
Hefei, China}

\title{\boldmath ATLAS Micromegas Performance Studies with LHC Run3 Data}

\author[a]{S. Francescato}
\collaboration[c]{on behalf of the ATLAS Muon Spectrometer group}

\affiliation[a]{Harvard University,\\
  Cambridge, Massachusetts, 02138}

\emailAdd{simone.francescato@cern.ch}

\abstract{After successfully completing Phase I upgrades during LHC Long Shutdown 2, the ATLAS detector is back in operation with several upgrades implemented. The most important and challenging upgrade is in the Muon Spectrometer, where the two inner forward muon stations have been replaced with the New Small Wheels (NSW) system. One of the two detector technologies used in the NSW are the resistive Micromegas (MM).
After massive construction, testing and installation work in ATLAS, the Micromegas are now fully operational in the experiment participating in the muon spectrometer tracking and trigger systems. A huge effort has gone into the operation of the new data acquisition system, as well as the implementation of a new processing chain within the muon software framework.
Tracking is performed with full consideration of the absolute alignment of each individual detector module by the ATLAS Muon Spectrometer optical alignment system. All the deviations from the nominal geometry of all the constituent elements of each MM module are accounted for through the modelling of the real chamber geometry reconstructed from the information of the construction databases.
After an overview of the strategies adopted for the simulations and reconstruction, the studies on the performance of the MM in LHC run-3 data taken from 2022 to 2024 will be reported.
}


\dedicated{\normalfont{Copyright 2025 CERN for the benefit of the ATLAS Collaboration. CC-BY-4.0 license.}}

\begin{document}
\maketitle
\flushbottom

\section{ATLAS New Small Wheels, design and status}
To address the increasingly challenging experimental conditions of the LHC \cite{LHC}, the ATLAS experiment \cite{ATLAS:2008xda} underwent significant upgrades during Long Shutdown 2 (LS2, 2019-2021) and has planned further improvements for Long Shutdown 3 (LS3, 2027-2030). The New Small Wheel (NSW), the largest ATLAS LS2 upgrade \cite{ATLAS:NSWTDR,ATLAS:2023dns}, was designed to handle the current Run 3 luminosity of $2 \cdot 10^{34}$ \si{cm^{-2}.s^{-1}} and the future HL-LHC \cite{HLLHC} peak luminosity of $5-7 \cdot 10^{34}$ \si{cm^{-2}.s^{-1}}.\\
This upgrade addressed two main challenges: improving Level-1 muon trigger efficiency to reject fake triggers caused by non-interaction-point particles and mitigating performance degradation in older detectors, particularly the Monitored Drift Tubes (MDTs), due to increased particle rates. The goals included maintaining a 20 GeV unprescaled muon trigger with high purity in the end-cap, confirming Big Wheel trigger coincidences with the NSW, and achieving high-resolution tracking at high momenta, aiming for 10\% muon momentum resolution at 1~TeV for rapidities $|\eta|$ up to 2.7.\\
The NSW replaced the innermost forward muon station detectors with Micro-Mesh Gaseous Structures (MicroMegas, MM) \cite{Alexopoulos:2018egk} and small-strip Thin Gas Chambers (sTGC). These technologies were designed to ensure $\mathcal{O}(100)$ \si{\mu m} track resolution, >95\% tracking efficiency for muons with transverse momentum >10~GeV, and resilience to high particle fluxes.\\
The upgrade involved constructing two detector wheels, one per ATLAS end-cap, each comprising 16 trapezoidal sectors. Every sector contains 16 detector planes arranged in 4 multi-layers, arranged in a sTGC-MM-MM-sTGC sequence, providing high redundancy for triggering and tracking. The system includes 357k sTGC readout channels and 2.05M MM readout channels. The VMM3a chip was adopted as the common front-end, while the new ATLAS data acquisition system FELIX was deployed on a large scale for the first time \cite{Iakovidis:2023ajz}.\\
This upgrade marks the first large-scale use of micropattern gaseous detectors in high-energy physics, with two 10 m diameter wheels equipped with 8 layers of MicroMegas covering a $160$~\si{m^2} active area, spanning $1.3<|\eta|<2.7$. ATLAS MicroMegas operate with a gas mixture of \ce{Ar}:\ce{CO2}:\ce{iC4H10} (93:5:2). Chambers feature a 5 \si{mm} drift gap at -300~V and a 120 \si{\mu m} amplification gap with readout planes at 500-520~V, with a grounded mesh. Spark protection is provided by resistive strips covering readout strips with a pitch of 425-450 \si{\mu m}, depending on the sector, ensuring precise spatial resolution. Of the 8 layers, 4 measure the precision $\eta$ coordinate, while 4 stereo layers tilted by $\pm 1.5^\circ$ provide the second coordinate. NSW has been part of ATLAS since the start of Run 3 in 2022, with the first year dedicated to commissioning. By 2024, ATLAS passed the 100 \si{fb^{-1}} at 13.6 \si{TeV} milestone, where NSW’s trigger rejection capabilities proved essential in minimizing dead time.

\section{NSW Micromegas performance}
\begin{figure}
    \centering
    \begin{subfigure}[b]{0.32\textwidth}
        \centering
        \includegraphics[width=\textwidth]{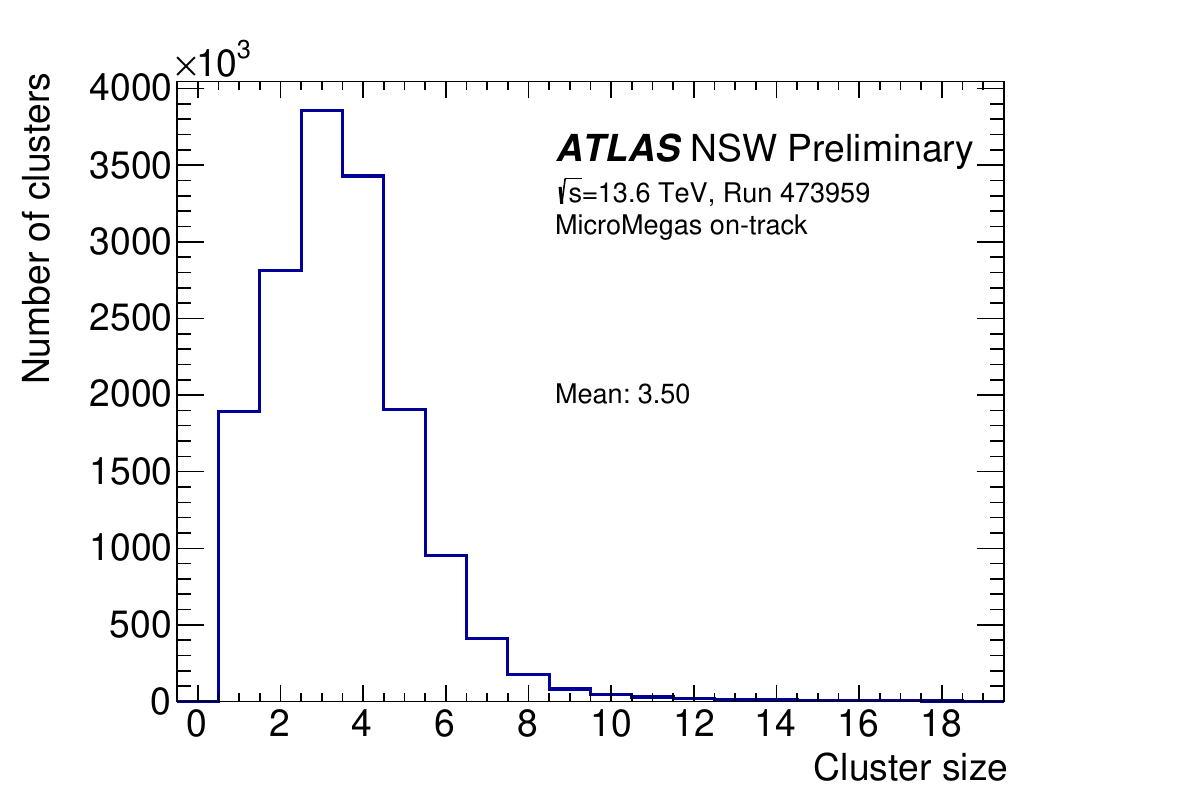}
        \caption{\label{fig:clusters:size}Cluster size}
    \end{subfigure}
    \hfill
    \begin{subfigure}[b]{0.32\textwidth}
        \centering
        \includegraphics[width=\textwidth]{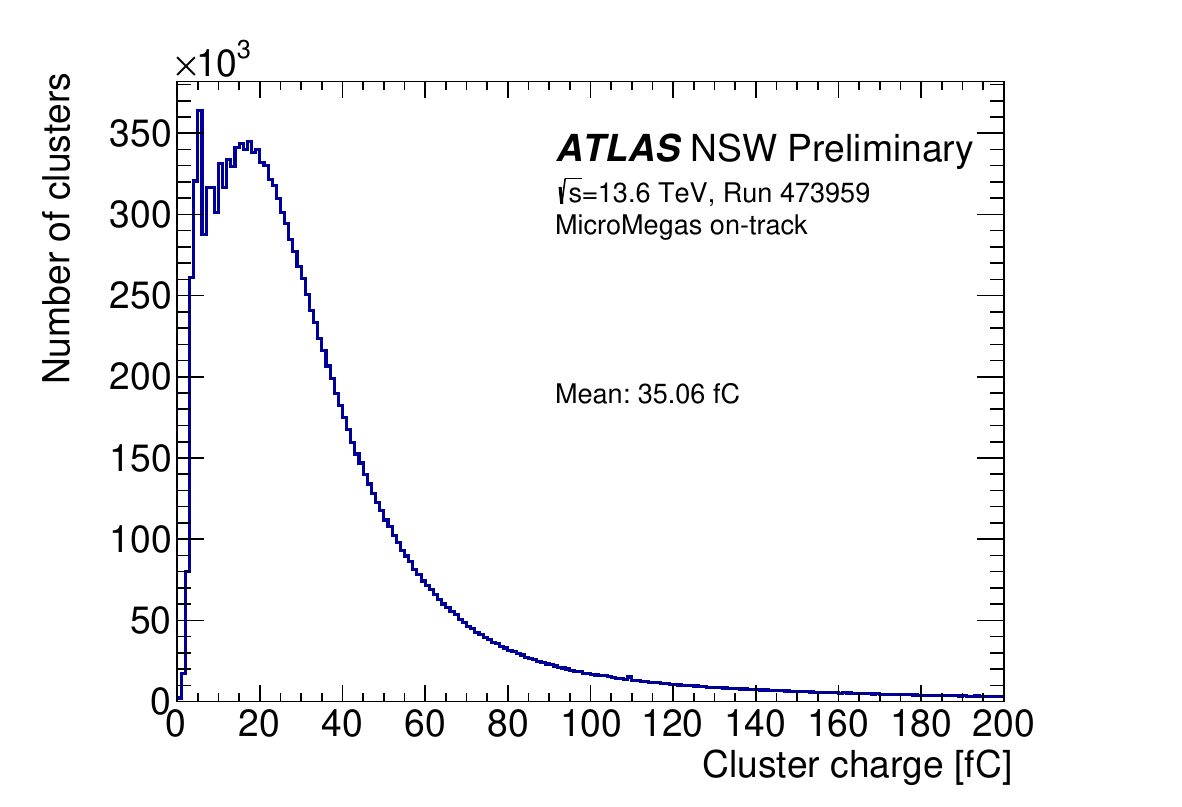}
        \caption{\label{fig:clusters:charge}Cluster charge}
    \end{subfigure}
    \hfill
    \begin{subfigure}{0.32\textwidth}
        \centering
        \includegraphics[width=\textwidth]{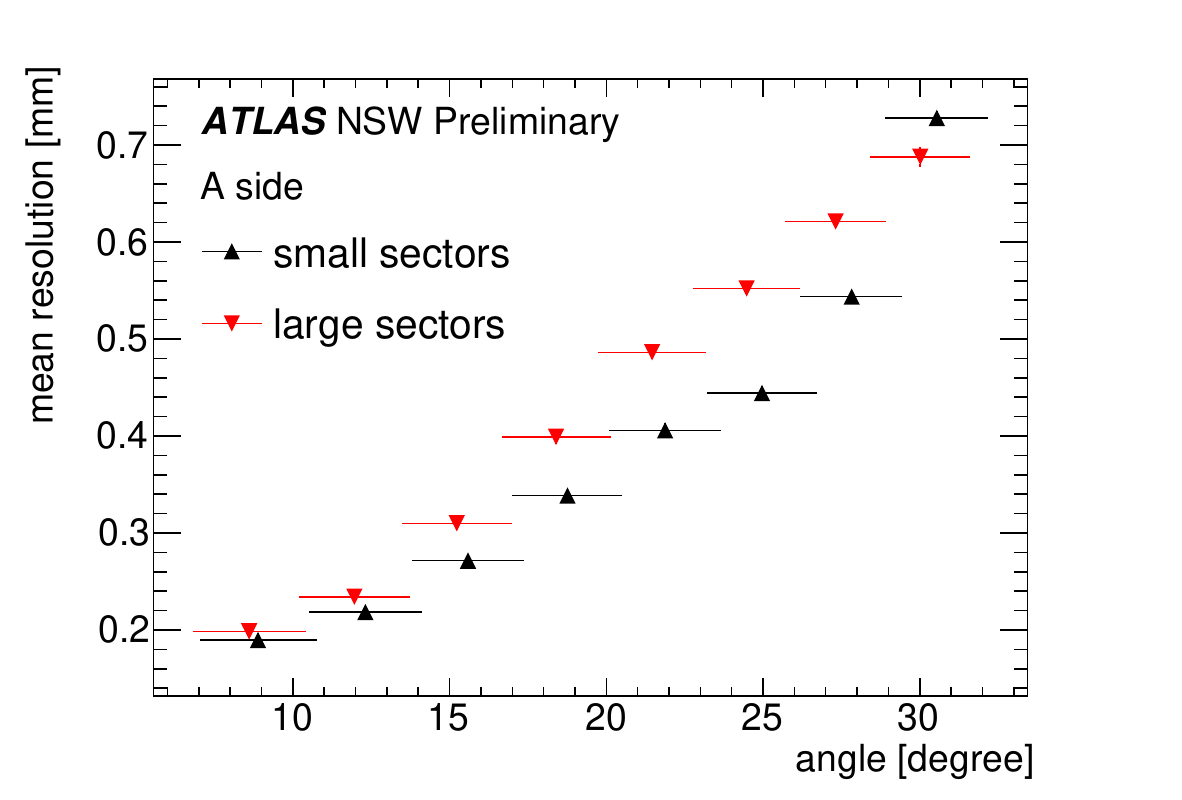}
        \caption{\label{fig:clusters:resol}Cluster resolution}
    \end{subfigure}
    \caption{\label{fig:clusters}MicroMegas cluster properties for clusters within $\pm 5$ \si{mm} of ATLAS Combined or Standalone muon tracks ($p_T > 15$ GeV) from 13.6 TeV pp collisions in 2024 \cite{Muon:NSW2024perf}. Position resolution is derived from neighboring layer cluster comparisons, corrected for track angle, avoiding alignment or as-built effects under study. Resolution is quoted as the 68\% confidence interval from a double Gaussian fit of residuals.}
\end{figure}

\begin{figure}
    \centering
    \includegraphics[width=0.91\textwidth, trim={5cm 0 0 4.5cm}, clip]{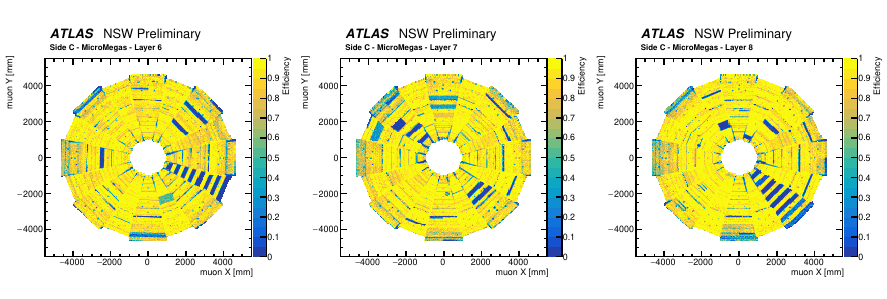}
    \caption{\label{fig:efficiency}Efficiency of selected Micromegas layers to reconstruct muon track positions in run 473959, evaluated using Combined or Standalone muon tracks ($p_T > 15$ GeV) from 13.6 TeV pp collisions \cite{Muon:NSW2024perf}. A layer is efficient if a cluster is within $\pm 5$ \si{mm} of the extrapolated track.}
\end{figure}

Since the start of Run 3, the NSW system has been fully integrated into the ATLAS data acquisition (DAQ) system and offline reconstruction environment. Currently, about 98.5\% of the MM chamber high-voltage channels are operational and collecting data.\\
One of the first NSW measurements was the cluster rate, which correlates with luminosity. Using 2023 data, the cluster rate was measured as a function of luminosity and distance from the LHC beam line, as shown in \cite{Muon:MM2024rates}. In the innermost NSW regions, rates reached 10 \si{kHz/cm^2} at luminosities of around $2.0-2.1 \cdot 10^{34}$ \si{cm^{-2}.s^{-1}}, with cavern backgrounds being the main contributor. Observed rates were ~30\% higher than expected from extrapolated Run 2 data, potentially due to shielding reductions in the Run 3 configuration, which is under study.\\
Initially, sTGC was designed as the primary fast trigger source for the NSW L1 Muon system. However, MM now plays a crucial role in ensuring high efficiency and rejection power. MM triggers use VMM ART hits\footnote{The VMM Address in Real Time (ART) output sends out the address of the channel with the earliest signal in every bunch crossing, scaling the channels down to 262k for trigger purposes. No charge or fine timing are provided.}, processed by FPGA-powered trigger boards. These require hits in 6 out of 8 layers pointing to the nominal IP, within an 8–10-strip road and an 8-bunch-crossing time window. After coincidences, a fast linear fit extracts MM trigger segment coordinates, which are merged with sTGC segments and sent to the ATLAS Endcap Muon trigger. By May 2024, the MM trigger was fully integrated, achieving >95\% efficiency across all NSW sectors and reducing fake rates further, as shown in \cite{Muon:MM2024trigger}. Fake MM trigger segments are now at sub-\% levels.\\
NSW is fully integrated into ATLAS simulation and reconstruction frameworks. Raw hits are mapped and clustered and used to build track segments across layers, contributing to global muon track fits. High-quality muon-associated NSW clusters enable detailed efficiency and performance studies. Currently, NSW cluster positions are extracted using the centroid method, based on charge-weighted strip averages.
Figure \ref{fig:clusters} summarizes significant cluster properties, with additional results in \cite{Muon:NSW2024perf}. Average cluster size is ~3 strips, increasing to 4 for highly inclined tracks ($30^\circ$ at the outer radius), with an average cluster charge of ~35 \si{fC}. The observed resolution ranges from 200–700 \si{\mu m}, while the design target is 100-150 \si{\mu m}, independently on the track angle. Alternative clusterization methods, including time-based algorithms, are being investigated to improve single-layer resolution and reduce $\eta$-dependent effects. Misalignment is a significant contribution to resolution limitations, with sagitta bias uncertainties of 80–100 \si{\mu m} in the NSW region \cite{Muon:NSW2024align}; given current detector efficiencies, with an average of about 6 planes per track, the combined resolution is indeed comparable alignment precision. Resolution differences between large and small sectors are instead driven by VMM setting differences. VMM thresholds tend to be higher for large modules, especially in the outermost NSW radii where strips are longer, since they are designed to maintain a consistent noise contribution. This effect is also sculpting the track angle dependency of the resolution.
Figure \ref{fig:efficiency} shows single-layer efficiencies for selected MM layers during a representative 2024 run. Also in this case, large and small sector differences are driven by the VMM threshold settings. Local inefficiencies stem from HV trips, DAQ instabilities or faulty LV boards. While run-dependent, efficiencies exceed 90\% in unaffected regions. Stable HV issues account for 2\% inefficiency, while dynamic LV/DAQ problems contribute 5\%. Some HV channels exhibit resistive behavior, affecting localized areas of a few \si{cm^2}. DAQ issues linked to VTRx chip problems were significantly mitigated between 2023 and 2024 \cite{Muon:NSW2023daq}. Overall, MM tracking efficiency - defined when 4 out of 8 layers have muon-associated clusters - is  above 95\%, stable throughout the year.

\section{Future prospects}
To prepare for HL-LHC, continuous irradiation studies of ATLAS Micromegas detectors are underway at the GIF++ facility. Chambers operate at 520 V and have stably sustained an accumulated charge of 0.29–0.34 \si{C/cm^2}, equivalent to 5–10 years of HL-LHC operation depending on strip location. While an expected resolution degradation with increased source intensity was observed, no ageing effects on efficiency or resolution were reported after returning to nominal Run 3 conditions, even after extensive irradiation. Further analyses, including detector gain and time resolution, are ongoing at GIF++, as detailed in \cite{Muon:MM2023gif}.


\end{document}